\documentclass[a4paper,fleqn,usenatbib]{mnras}

\usepackage{newtxtext,newtxmath}

\usepackage[T1]{fontenc}
\usepackage{ae,aecompl}
\usepackage{bm}
\usepackage{xcolor}
\usepackage{scalerel}
\usepackage{tikz}
\usetikzlibrary{svg.path}

\definecolor{orcidlogocol}{HTML}{A6CE39}
\tikzset{
  orcidlogo/.pic={
    \fill[orcidlogocol] svg{M256,128c0,70.7-57.3,128-128,128C57.3,256,0,198.7,0,128C0,57.3,57.3,0,128,0C198.7,0,256,57.3,256,128z};
    \fill[white] svg{M86.3,186.2H70.9V79.1h15.4v48.4V186.2z}
                 svg{M108.9,79.1h41.6c39.6,0,57,28.3,57,53.6c0,27.5-21.5,53.6-56.8,53.6h-41.8V79.1z M124.3,172.4h24.5c34.9,0,42.9-26.5,42.9-39.7c0-21.5-13.7-39.7-43.7-39.7h-23.7V172.4z}
                 svg{M88.7,56.8c0,5.5-4.5,10.1-10.1,10.1c-5.6,0-10.1-4.6-10.1-10.1c0-5.6,4.5-10.1,10.1-10.1C84.2,46.7,88.7,51.3,88.7,56.8z};
  }
}

\newcommand\orcidicon[1]{\href{https://orcid.org/#1}{\mbox{\scalerel*{
\begin{tikzpicture}[yscale=-1,transform shape]
\pic{orcidlogo};
\end{tikzpicture}
}{|}}}}

\usepackage{float}
\usepackage[export]{adjustbox}
\usepackage{graphicx}	

\usepackage{amsmath}	
\usepackage{amssymb}	
\usepackage{multirow}
\usepackage{multicol}
\usepackage{verbatim}   
\usepackage{threeparttable}
\usepackage{upgreek}
\usepackage{rotating}

\usepackage{subcaption}
\usepackage{comment}
\usepackage{longtable,lscape}
\usepackage{caption}
\usepackage{threeparttable}
\newcommand{\FT}[1]{}

\definecolor{MZ}{RGB}{255,128,0}

\title[Breaking MSD using weak lensing observations]{Breaking the mass-sheet degeneracy in strong lensing mass modeling with weak lensing observations}

\author[Khadka et al.]{
Narayan Khadka,$^{{1}}$\thanks{E-mail: narayan.khadka@stonybrook.edu}
Simon Birrer,$^{1}$\thanks{E-mail: simon.birrer@stonybrook.edu}, Alexie Leauthaud,$^{2}$ Holden Nix,$^{2}$\\
$^{1}$Department of Physics and Astronomy, Stony Brook University, Stony Brook, NY 11794, USA\\
$^{2}$Department of Astronomy and Astrophysics, University of California, Santa Cruz, 1156 High Street, Santa Cruz, CA 95064 USA\\
}
\date{Accepted XXX. Received YYY; in original form ZZZ}

\pubyear{2024}

\begin{document}
\label{firstpage}
\pagerange{\pageref{firstpage}--\pageref{lastpage}}
\maketitle

\begin{abstract}
The Hubble constant ($H_0$), a crucial parameter in cosmology, quantifies the expansion rate of the universe so its precise measurement is important to understand the fundamental dynamics of our evolving universe. One of the major limitations of measuring $H_0$ using time-delay cosmography is the presence of the mass-sheet degeneracy (MSD) in the lens mass modeling. We propose and quantitatively assess the use of galaxy-galaxy shear measurements to break the MSD in the strong lensing mass modeling. We use stacked galaxy-galaxy lensing profiles and corresponding covariance matrices from Huang et al.\ (2022) to constrain the MSD in lens mass modeling with a highly flexible mass profile. Our analyses show that if ideally all galaxy-galaxy lensing measurements from the Hyper Suprime-Cam (HSC) survey can be used to constrain the MSD, we can achieve $\sim 10\%$ precision on the MSD constraint. We forecast that galaxy-galaxy lensing measurements from LSST-like surveys can in general constrain the MSD with $\sim 1-3\%$ precision. Furthermore, if we push weak lensing measurements to a lower angular scale of $\sim 0.04 \rm Mpc$, a survey like LSST can provide $\sim 1\%$ precision on the MSD constraint, enabling a measurement of $H_0$ at the 1\% level. We demonstrate that galaxy-galaxy weak lensing can robustly constrain the MSD independent of stellar kinematics of the deflector, with wide-field survey data alone. 

\end{abstract}

\begin{keywords}
\textit{(cosmology:)} observations -- \textit{(gravitational lensing:)} strong -- \textit{(gravitational lensing:)} weak -- \textit{(galaxies:)} haloes 
\end{keywords}



\section{Introduction}
\label{sec:Introduction}
In the last few decades, observational cosmology has acquired a considerable amount of data that demands dark energy as a major contributor in the cosmological energy budget. The standard model of cosmology i.e. spatially-flat $\Lambda$CDM model \citep{Peebles1984} can approximate remarkably well the observational data when attributing $\sim 70\%$ of the energy budget to the cosmological constant $(\Lambda)$, $\sim 25\%$ to the cold dark matter, and the remaining $\sim 5\%$ to the baryonic matter \citep[see, e.g.][]{Scolnicetal2018, Yuetal2018, PlanckCollaboration2020,eBOSSCollaboration2021}. However, some of the cosmological parameters in the spatially-flat $\Lambda$CDM model seem to be mutually incompatible while considering different observational data \citep[see, e.g.][]{DiValentinoetal2021b, PerivolaropoulosSkara2021, Abdallaetal2022}.

One such incompatibility is in the measurement of the Hubble constant ($H_0$). Values of $H_0$ inferred using early-universe measurements compared with directly measured values using late-universe probes are significantly discrepant \citep{PlanckCollaboration2020, Riess2022}. 
%
These values are in more than $4\sigma$ tension within a spatially-flat $\Lambda$CDM model, known as the "Hubble tension". The discrepancy can be either due to (1) implications from new physics beyond the standard model or (2) caused by unaccounted-for uncertainties in one or several measurements.

Developing multiple independent and precise  cosmological probes to measure the absolute scales of the universe is essential to shed conclusive light on the Hubble tension.
In particular, measurements of $H_0$ using a probe independent of early-universe physics and the local distance ladder has the possibility of conclusively solving the $H_0$ tension. One such probe is multiply-imaged gravitational lensed time delays, also known as Time-delay Cosmography \citep[TDC;][]{Refsdal:1964}. 
A continuous scientific effort has developed TDC as a major probe of $H_0$ using lensed quasars \citep[e.g.,][and references therein]{Vanderriest1989, Keeton1997, Schechter1997, 2003Kochanek, 2003Koopmans, Saha2006, Oguri2007, Suyu2010, Sereno2014, Rathna2015, Birrer2016, Wong2017, Birrer2019, Rusu2019, Chen2019, Shajib2020, Birrer2020}.
We also refer to e.g.,  
\citet{Treu:2022, Birrer_review} for recent reviews. 

The H0 Lenses in COSMOGRAIL’s Wellspring (H0LiCOW) \citep{Suyu2010, suyu2017} and Time-Delay COSMOgraphy (TDCOSMO) \citep{Millonetal2020, Birrer2020} collaborations analyzed seven lensed quasars and assuming specific parameteric forms of the radial density profile of the deflector galaxy achieved a 2.4 per cent precision measurement with $H_0 = 73.3^{+1.7}_{-1.8}$ $\rm km$ $\rm s^{-1}$ $\rm Mpc^{-1}$ \citep{Wongetal2020}.
This work demonstrates that, provided sufficient knowledge of the radial density profiles of the deflectors, the road to a decisive $1\%$ measurement of $H_0$ is wide open with an anticipated sample of $>100$ lensed quasars \citep{OguriMarshall2010} and an already substantial size of recently discovered lensed quasars \citep{Schmidt:2023}.
In addition to lensed quasars, the discovery of the first multiply-imaged supernovae (SN) \citep{Kelly:2015} has enabled to measure $H_0$ with lensed SNe \citep{Vega-Ferrero:2018, Kelly:2023}. With the Vera C. Rubin Observatory, we anticipate the discovery of $\sim 44$ lensed SNe per year of the survey with $\sim 10$ lensed SNe being particularly promising for a precise time-delay measurement \citep{Arendse:2023}.


The primary systematic uncertainty of the $H_0$ measurement with TDC is due to the mass-sheet degeneracy \citep[MSD;][]{Falco1985}, which originates from the mathematical mass-sheet transform (MST) in the lens equation.
The MST is a mathematical degeneracy that leaves
the lensing observables unchanged, while rescaling the absolute time delay, and thus the inferred value of $H_0$.
The MSD has sparked a long-lasting discussion in the literature \citep{Gorenstein:1988, Kochanek:2002, Read:2007, Schneider:2013, Schneider:2014, Coles:2014, Xu:2016, Birrer2016, Unruh:2017, Sonnenfeld:2018, Kochanek:2020, Blumetal2020}. It is absolutely vital to constrain the radial mass density profile degeneracy related to the MST for an accurate and precise measurement of $H_0$.

Lensing-independent tracers of the gravitational potential of the deflector galaxy, such as stellar kinematics, can break this inherent degeneracy \citep{Grogin1996, Romanowsky1999, Treu2002, Barnabeetal2012, Barnabe2011}. This was the approach chosen for the recent $H_0$ measurement by the TDCOSMO collaboration \citep{Birrer2020}. \cite{Birrer2020} used a maximally degenerate radial lensing profile and only used the stellar kinematic measurements to constrain the MSD with a hierarchical Bayesian approach.
Using seven TDCOSMO time-delay lenses alone, they measured $H_0 = 74.5^{+5.6}_{-6.1}$ $\rm km$ $\rm s^{-1}$ $\rm Mpc^{-1}$, while combining the TDCOSMO lenses with 33 Sloan Lens ACS sample \citep{Bolton:2006, Shajib_slacs:2021} provides $H_0 = 67.4^{+4.1}_{-3.2}$ $\rm km$ $\rm s^{-1}$ $\rm Mpc^{-1}$ \citep{Birrer2020}.

One limitation of using stellar kinematics to constrain the radial density profile becomes apparent: the precision in the kinematics measurements used by \cite{Birrer2020} were not sufficient for a precise measurement of $H_0$.
To measure a precise velocity dispersion, expensive follow-up observations by 8-meter class ground-based telescopes or space-based facilities are required and a major effort is underway to tighten these uncertainties \citep[e.g.,][]{Shajib:2023} with spatially resolved kinematics measurements and a path towards a $1\%$ precision is laid out by \cite{BirrerTreu:2021} using a set of $\sim 40$ lensed quasars with spatially resolved kinematics from JWST.
Beyond the exquisite data required, there is the challenge for an accurate interpretation and modeling of the measured stellar spectra. The measurement of the dispersion in the spectra requires knowledge and assumptions on the stellar template spectra, and the dynamical modeling requires knowledge about the velocity anisotropy distribution of the stars throughout the phase space, known as the mass-anisotropy degeneracy \citep{Merritt1987}. While the mass-anisotropy degeneracy can be contained with spatially resolved measurements, other secondary assumptions, like the axis-symmetry, the intrinsic triaxiality of the mass and light distribution, rotational components and the assumption of equilibrium need to be further investigated.

Hence, the development of independent and complementary methodology to break the MSD are highly desired. Another method to break the MSD that was proposed is using absolute lensing magnification constraints from lensed type-Ia SNe \citep{Kolatt:1998, OguriKawano:2003, Foxley-Marrable:2018, BirrerSNe:2022}. This method requires extensive and prompt spectroscopic and imaging follow-up observations, a well standardizable magnitude of SNe to higher redshift, as well as the precise mitigation of the micro-lensing impact on the lensing magnification \citep[see e.g.,][]{Foxley-Marrable:2018}.

In this paper, we put forward and quantitatively assess a new method to constrain the MSD-related uncertainty in strong lens mass modeling using galaxy shape weak gravitational lensing measurements. We examine the use of stacked galaxy shear/lensing profiles of self-similar galaxies to constrain the MSD for individual strong lensing deflectors. The close statistical match of a larger population of deflector for the weak lensing analysis with a strong lensing sample is a requirement for this method to work.
Combined analyses of strong and weak gravitational lensing are a common practice in galaxy cluster analyses \citep[e.g.,][and references therein]{Bradac2004, Bradac:2005, Cacciato:2006, Diego:2007, Merten:2009} and have also successfully been applied to galaxy-scale gravitational lenses to constrain the halo mass and the shape of the dark matter profile
\citep[e.g.,][and references therein]{Gavazzi:2007, Sonnenfeld:2018}. In this work we present a methodology to constrain the MSD-effect relevant for the measurement of the Hubble constant.

\begin{figure*}
    \includegraphics[width=\linewidth,right]{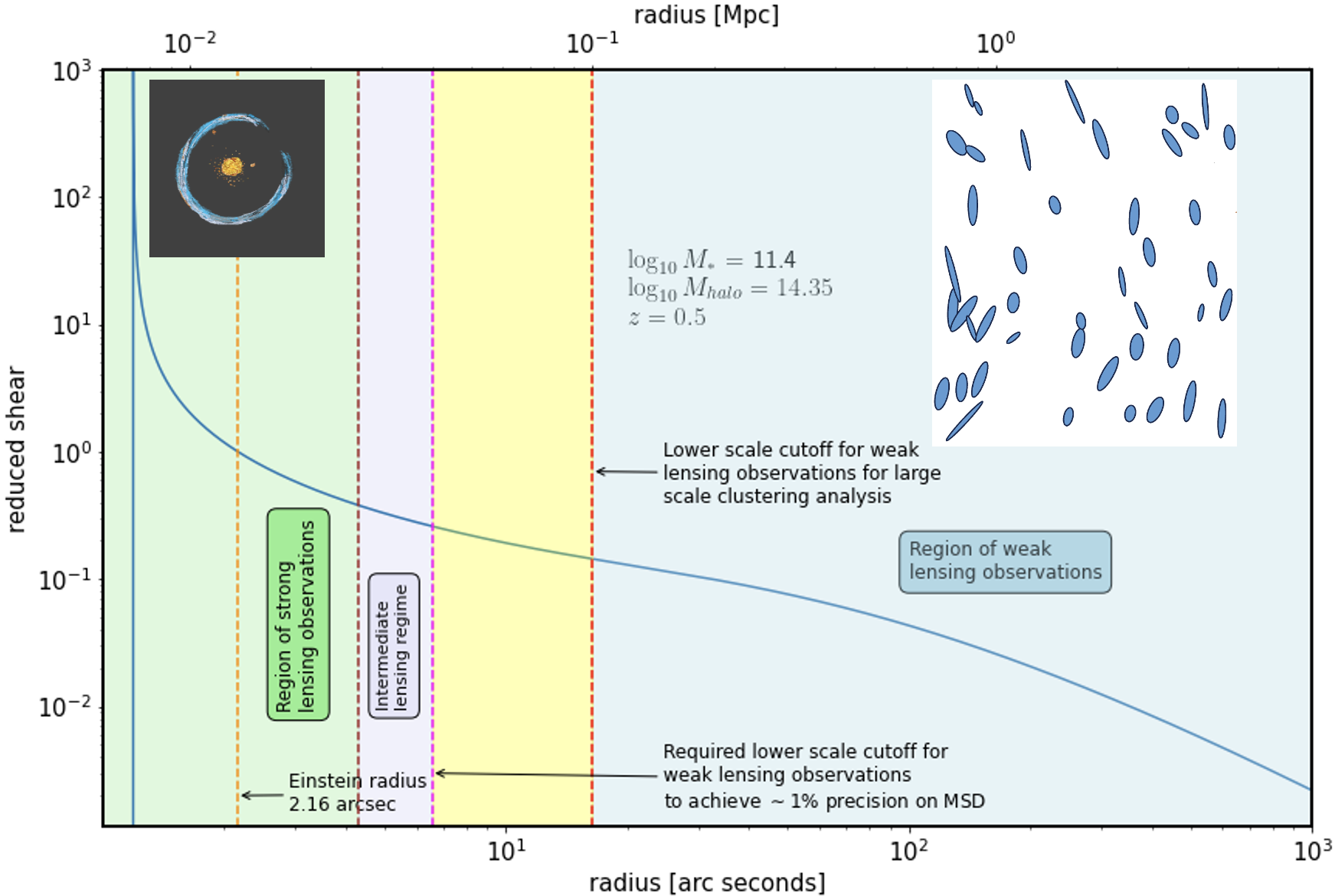}\par
\caption{Reduced shear profile of a lens with an illustration of strong and weak gravitational lensing at $z = 0.5$ regimes. The light blue shaded region in the right hand side corresponds to the weak lensing observations where weak lensing shape measurements are generally available. The vertical red dashed line represents the lower limit of the weak lensing measurements for large scale clustering analyses. The light green shaded region in the left corresponds to the strong lensing regime. The orange dashed line represents the Einstein radius of the lens and brown dashed line represents twice the Einstein radius which in general is the upper limit for strong lensing observations. The light lavender shaded region corresponds to the intermediate lensing regime. The vertical magenta dashed line is the limit where weak lensing measurements need to reach in order to achieve $\sim 1\%$ precision on the MSD. The yellow shaded region is the gap needed to fulfill by weak lensing data to achieve $\sim 1\%$ precision on the MSD constraint. $M_{\rm halo}$ and $M_{*}$ in the light blue shaded region are halo mass and stellar mass respectively.
}
\label{fig:illustration}
\end{figure*}

The mathematical form of the MST is an infinitely extending sheet of constant projected mass density. There is no physically viable interpretation that results in a pure MST. Any density perturbation eventually has to decline to the cosmological background density at some radial scale.
There exist a variety of physically valid scenarios where the excess density goes to zero at large angular radii that approximate to high precision the effect of an MST on the scale of strong lensing observables \citep[see e.g.,][]{Blumetal2020, Birrer2020}. Any of these valid approximations to an MST will lead to imprints on the shear profile at some angular radius far beyond the Einstein radius of a strong lens. If we have a continuous, or sufficiently continuous, radial measurement of the shear profile from the Einstein radius out to scales where we confidently can state that there should not be excess surface mass density to be present, we can break all components of the MSD relevant for TDC. 
An illustration of the strong and weak lensing regimes are shown in Fig.\ \ref{fig:illustration}.

Our proposed method brings two key advantages compared to other existing or proposed methods:
(1) Weak lensing directly measures signal related to the projected radial density profile, the quantity essential for TDC and hence systematics attributed to secondary effects and assumptions, such as de-projection or micro-lensing, are avoided.
(2) The required data, and most likely even the derived shape measurements and photometric redshifts, are either already existing or highly anticipated as key drivers of current and near-future wide-field ground- and space-based surveys, such as the Vera C. Rubin Observatory, Euclid, or the Nancy Grace Roman Observatory \citep{LSST2019, Euclid2022, Roman2023}. No targeted follow-up efforts beyond these survey data products are required to apply this method to break the MSD.

In this paper, aside from proposing galaxy-galaxy lensing to break the MSD, we introduce a generalized parametric form of approximate mass sheets that exert all possible degrees of flexibility in the density profile at outer radii while conserving strong lensing constraint. We then use current available galaxy-galaxy weak lensing data from HSC to constrain the general approximate MSD profile. We then use the current data vector as a baseline to forecast constraints from anticipated data from the Vera C. Rubin Observatory. 
We assess that with the future data set we can constrain the MSD-component in the radial density profile to $\sim 1\%$ on its impact on $H_0$. Given the anticipated precision with data readily to be available and the limited systematics introduced, this method integrated into a TDC analysis is able to decisively shed light on the Hubble tension.




This paper is organized as follows. In Section \ref{sec:methods} we describe the methodology we develop. In Section \ref{sec:data} we describe the data we use. In Section \ref{sec:results} we report our results and in Section \ref{sec:discussion} we discuss some challenges with weak lensing measurements and strong lensing selection bias. We conclude in Section \ref{sec:conclusion}. 

\section{Methods}
\label{sec:methods}
The strong lensing phenomena can be described by the lens equation,
\begin{equation}
\label{eq:lens_equ}
    \vec\beta = \vec\theta - \vec\alpha (\vec\theta),
\end{equation}
where, $\vec\theta$ is an image position, $\vec\beta$ is a source position, and $\vec \alpha$ is a deflection angle of the source light because of the lensing. The deflection angle $\vec\alpha$ is related with the lensing potential $\phi(\theta)$ of a lens by the equation,
\begin{equation}
\label{eq:lensing_potential}
    \nabla\phi(\vec\theta) = \vec\alpha (\vec\theta),
\end{equation}

When a variable source is strongly lensed by a massive object, the intrinsic variability pattern of the source can be observed in each image in the lens system. Based on these patterns, one can measure the time delay between each pair of images due to the different path. The time delay between two images at $\vec\theta_A$ and $\vec\theta_B$ of the same source at $\vec\beta$ is given by,
\begin{equation}
\label{eq:time_delay}
    \Delta t_{AB} = \frac{D_{\Delta t}}{c} [\zeta(\vec\theta_A, \vec\beta) - \zeta(\vec\theta_B, \vec\beta)],
\end{equation}
where, $c$ is the speed of light, $\zeta$ is the Fermat potential \citep{Schneider1985, Blandford1986} and is given by,
\begin{equation}
\label{eq:Format_potential}
    \zeta(\vec\theta) = \left[\frac{(\vec\theta - \vec\beta)^2}{2} - \phi(\vec\theta)\right],
\end{equation}
and $D_{\Delta t}$ is the time-delay distance and is given by,
\begin{equation}
\label{eq:time_delay_distance_pred}
    D_{\Delta t} = (1 + z_{\rm l})\frac{D_{\rm os} D_{\rm ol}}{D_{\rm ls}},
\end{equation}
where $z_{\rm l}$ is the redshift of the lens. $D_{\rm os}$, $D_{\rm ol}$, and $D_{\rm ls}$ are the angular diameter distance between observer and source, observer and lens, and lens and source respectively.

From eq.\ \ref{eq:time_delay}, the measured time delay and constraints on Fermat potential can be used to determine the time-delay distance,
\begin{equation}
\label{eq:time_delay_distance_measure}
    D_{\Delta t} = \frac{c \Delta t_{AB}}{\Delta\zeta_{AB}}.
\end{equation}
The predicted time-delay distance using eq.\ \ref{eq:time_delay_distance_pred} can be compared with the measured time-delay distance from eq.\ \ref{eq:time_delay_distance_measure} and this allows us to constrain $H_0$. However, the Fermat potential is impacted by the MSD. This is a mathematical degeneracy which results in exactly the same image observables with the scaled source configurations. Mathematically, this is a transformation of the lens equation and can be expressed as

\begin{equation}
\label{eq:MST}
    \lambda \vec\beta = \vec\theta - \lambda \vec\alpha (\vec\theta) + (1-\lambda) \vec\theta,
\end{equation}

here, $\vec\theta$ is a preserved quantity in this equation. $\lambda$ is a transformation factor. With this transformation, the Hubble constant transforms as,
\begin{equation}
\label{eq:H0scale}
    H_{0\lambda} = \lambda H_0.
\end{equation}
Therefore, MSD is the major source of systematic error in $H_0$ measurement. So, an accurate constraint on $\lambda$ is necessary for the precise measurement of the Hubble constant.

In the convergence $(\kappa)$ field, this transformation can be expressed as
\begin{equation}
\label{eq:MSTC}
    \kappa_{\lambda} = \lambda \kappa (\vec\theta) + (1-\lambda),
\end{equation}
here, $\kappa_{\lambda}$ is the transformed convergence, $\kappa (\theta)$ is the original convergence. $\vec\theta$ is the angular position from the center of the mass-sheet. The last term, $(1-\lambda)$, provides convergence of the mass sheet in the lens model.

The convergence associated with the internal mass sheet transformation should drop to zero at infinity, i.e.\, $\lim_{\vec\theta\to\infty} \kappa(\vec\theta) = 0$. Therefore, the internal mass sheet transformation can be modeled as an approximate mass sheet transformation as a function of mass-sheet convergence ($\kappa_c$) \citep{Blumetal2020}. This can be expressed as,

\begin{equation}
\label{eq:amst}
    \kappa_{\lambda_c} = \lambda_c \kappa (\vec\theta) + (1-\lambda_c)\kappa_c (R_c, \vec\theta),
\end{equation}
\begin{figure*}
    \includegraphics[width=\linewidth,right]{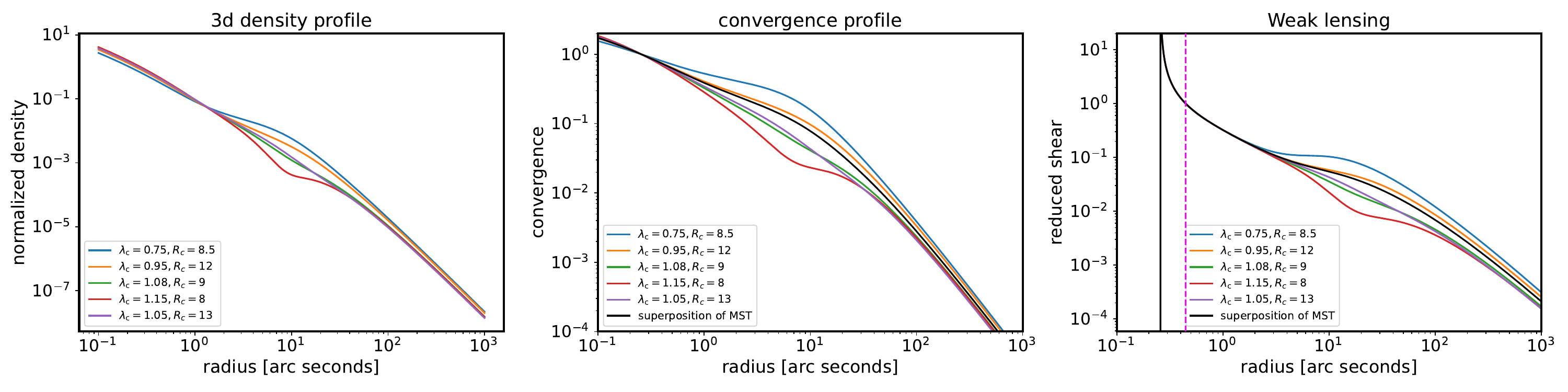}\par
\caption{Superposition of five different mass-sheet transformations. Black curves in the middle and right panels are the resultant convergence and reduced shear from the superposition of different mass-sheets which are given in other colors. $\lambda_{\rm c}$ and $R_{\rm c}$ are the MST factor and the core radius of a mass-sheet corresponding to each profile, and these quantities are described in eq.\ \ref{eq:conv}. The left, middle, and right panels show the three-dimensional density profile, convergence profile, and weak lensing profile of the halo, respectively. The vertical magenta dashed line in right panel represents the  Einstein radius of the halo and its value is 0.44 arcsec. Different profiles shown in each panel are indistinguishable for the strong lensing imaging data. However, weak lensing reduced shear measurements can distinguish these profiles.}
\label{fig:smst}
\end{figure*}
here, $R_c$ is a core radius of the mass-sheet and $\kappa_c$ is given by,

\label{eq:kc}
    \begin{equation}
    \kappa_c = \frac{R_c^2}{R_c^2 + \vec\theta^2}.
\end{equation}
As a physical requirement mentioned above, this convergence term goes to zero as $\vec\theta$ goes to infinity.

\subsection{Proposed model for the approximate MST}
\label{sec:proposed_model}
Instead of relying on a fixed mass-sheet transformation, we model the MST by superimposing numerous potential approximate mass-sheet transformations. The superposition of approximate mass sheet transformations is also an approximate MST. So, this model yields a physically plausible mass-sheet transformation as mentioned above i.e $\kappa_c \to 0$ as $\vec\theta \to \infty$ which can be seen from eq.\ \ref{eq:conv} below. This is a novel approach to deal with the MST which provides more flexible and realistic representation of the convergence profile. The superposition of possible mass-sheet can be expressed as
\begin{equation}
\label{eq:conv}
    \kappa_{\lambda_c}(\vec\theta, \{\lambda_{c, i}\}, \{R_{c, i}\}) = (1-\sum_i \kappa_{c, i}) \kappa (\vec\theta) + \sum_i \kappa_{c, i} \frac{R_{c, i}^2}{R_{c, i}^2 + \vec\theta^2},
\end{equation}
where, $\kappa_{c, i} = (1-\lambda_{c, i})$ is the convergence of the $i^{th}$ mass-sheet and $\lambda_{c, i}$ is the transformation factor for the $i^{th}$ approximate mass-sheet transformation. $\kappa (\vec\theta)$ is a predicted convergence using a model or a combination of models. In this paper we use a combination of NFW and Hernquist profile to predict $\kappa(\vec\theta)$. $\lambda_{c, i}$ and $R_{c, i}$ are free parameters to be determined using external data. One can also choose only $\lambda_{c, i}$ to be a free parameter and can fix $R_{c, i}$ to specific values with in the profile. In this equation, the total MST factor is given by,
\begin{equation}
\label{eq:lambda_tot}
    \lambda_{\rm tot} = (1-\sum_i \kappa_{c, i}),
\end{equation}
and this is the quantity that we are interested in and want to constrain using external data. $\lambda_{c, i}$ are correlated with each other so the constraints on each individual $\lambda_{c, i}$ could be loose but the constraint on $\lambda_{\rm tot}$ could be much narrower compared to constraints on each individual MST.

The implication of eq. (\ref{eq:conv}) can be visualized in Fig.\ \ref{fig:smst}. This figure shows that this profile can have a very flexible shape because superposition of different MSTs can have better control in the shape of the profile.

\subsection{Theoretical galaxy-galaxy weak lensing profile}
\label{sec:gg_lensing}
The tangential shear profile corresponding to convergence in eq.\ \ref{eq:conv} is given by,
\begin{equation}
\label{eq:shear}
\begin{aligned}
    \gamma(\vec\theta, \{\lambda_{c, i}\}, \{R_{c, i}\}) = \frac{2}{|\vec\theta|^2}\int_{0}^{\vec\theta} \kappa_{\lambda_{c}}(\vec\theta^{\prime}, \{\lambda_{c, i}\}, \{R_{c, i}\}) \vec\theta^{\prime} d\vec\theta^{\prime}\\ - \kappa_{\lambda_{c}}(\vec\theta, \{\lambda_{c, i}\}, \{R_{c, i}\}),
\end{aligned}
\end{equation}
and the excess surface mass density associated with this is given by,
\begin{equation}
\label{eq:dsigma}
    \Delta \Sigma_{\rm th}(\vec\theta, \{\lambda_{c, i}\}, \{R_{c, i}\}, z_{\rm l}, z_{\rm s}) = \gamma(\vec\theta, \{\lambda_{c, i}\}, \{R_{c, i}\})\times\Sigma_{\rm crit}(z_{\rm l}, z_{\rm s}),
\end{equation}
where $z_{\rm s}$ is the source redshift and $\Sigma_{\rm crit}$ represents the critical surface mass density of a lens, defined as,
\begin{equation}
\label{eq:sigma_crit}
    \Sigma_{\rm crit} = \frac{c^2}{4\pi G} \frac{D_{\rm os}}{D_{\rm ol}D_{\rm ls}},
\end{equation}
here, $G$ is the Newton's gravitational constant. 

\subsection{Observed galaxy-galaxy weak lensing profile}
\label{sec:observed_lp}
Distant background galaxies, far beyond the Einstein radius, experience weak gravitational lensing from the lens galaxy. This lensing effect manifests as shape distortions in the background galaxies. These distortions provide a rich statistical basis for measuring the shear profile and subsequently determining the excess surface mass density, referred to as the lensing profile, of a lens galaxy. Therefore, a fundamental observable in weak lensing scenarios is the shape measurements of background galaxies \citep{Mandelbaumetal2018, Mandelbaumetal2017}.

One can bin all the shear measurements of background galaxies in a radial bin of a lens galaxy and convert shear measurements of a lens and source pair to $\Delta\Sigma$ by multiplying with critical surface mass density of corresponding lens and source pair. Then, weighted mean of shear within each bin provide an observed shear profile for a lens galaxy. Mathematically, the weighted mean of excess surface mass density with in each bin is given by,
\begin{equation}
\label{eq:delta_sigma}
    \Delta\Sigma_{\rm obs}(z_{\rm l}, z_{\rm s}) = \frac{\sum_{\rm ls} w_{\rm ls}\gamma^{\rm ls}_t \Sigma^{\rm ls}_{\rm crit}(z_{\rm l}, z_{\rm s})}{\sum_{\rm ls} w_{\rm ls}},
\end{equation}
where,
\begin{equation}
\label{eq:tan-shear}
    \gamma^{\rm ls}_t = -[\gamma_1 \cos(2\phi) + \gamma_2 \sin(2\phi)].
\end{equation}
Here, $w_{\rm ls}$ is a weight associated with tangential shear $(\gamma^{\rm ls}_t)$ of each lens and source pair. $\phi$ is the angle of the vector connecting lens and the source, and ($\gamma_1$ and $\gamma_2$) are the measured shear component of the source. Observed random signal can be subtracted from the eq.\ \ref{eq:delta_sigma} to reduce random noise.

\subsection{Likelihood}
One can predict the lensing profile of a lens ($\Delta \Sigma$) using eq. \ \ref{eq:dsigma} along with eqs. \ \ref{eq:shear} and \ref{eq:sigma_crit} as a function of mass-sheet transformation factors ($\{\lambda_{c, i}\}$) and core radii $\{R_{c, i}\}$. Then, we can compare these predicted lensing profiles with the observed lensing profile by using the $\log$ likelihood function,
\begin{equation}
\label{eq:llike}
\begin{aligned}
    \ln({\rm LF}) = -\frac{1}{2} [\Delta\Sigma_{\rm obs}(z_{\rm l}, z_{\rm s}) - \Delta\Sigma_{\rm th}(\vec\theta, \{\lambda_{c, i}\}, \{R_{c, i}\}, z_{\rm l}, z_{\rm s})]^T \textbf{C}^{-1}\\
    [\Delta\Sigma_{\rm obs}(z_{\rm l}, z_{\rm s}) -\Delta\Sigma_{\rm th}(\vec\theta, \{\lambda_{c, i}\}, \{R_{c, i}\}, z_{\rm l}, z_{\rm s})],
\end{aligned}
\end{equation}
where $\Delta\Sigma_{\rm obs}(z_{\rm l}, z_{\rm s})$ and $\Delta\Sigma_{\rm th}(\vec\theta, \{\lambda_{c, i}\}, \{R_{c, i}\}, z_{\rm l}, z_{\rm s})$ are the measured and theoretical lensing profiles, respectively. \textbf{C} is the covariance matrix of an observed lensing profile.

The likelihood analysis is performed using the Markov chain Monte Carlo (MCMC) method, which is implemented in the \texttt{emcee} package \citep{emcee}. Lens mass modeling is conducted using \texttt{lenstronomy} \citep{lenstronomy, lenstronomyII}.

\begin{figure*}
    \includegraphics[width=\linewidth]{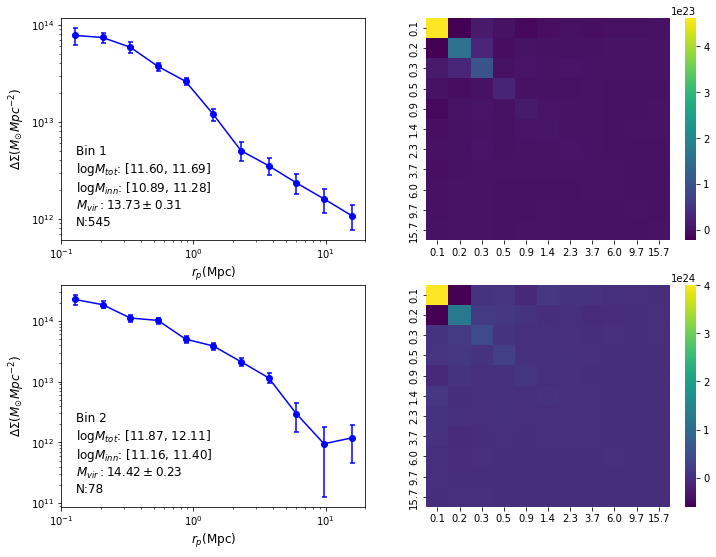}\par
\caption{Stacked galaxy-galaxy lensing profiles and corresponding covariance matrices from the HSC data for two different mass bins \citep{Huang2022}. Left column shows lensing profiles and the right column shows corresponding covariance matrices. These two bins are representative bins for the high and low mass halos in the massive galaxy sample of the HSC data. Covariance matrices of these two profiles have been used to constrain the MSD in a mock lensing profile in Sections \ref{re:obs_results} and \ref{sec:forecast_results}.}
\label{fig:obser_data}
\end{figure*}

\section{Data}
\label{sec:data}
\cite{Huang2022} has identified a massive lens galaxy sample from $\sim 137$ deg$^2$ of the deep-optical images from the Hyper Suprime-Cam (HSC) survey \citep{Mandelbaumetal2017, Mandelbaumetal2018} which span redshift range $0.19 < z < 0.51$. They have computed 27,207 individual galaxy-galaxy lensing profiles. Signal to noise ratio for each individual profile is low so one cannot use individual profiles for any cosmological application. To increase the signal-to-noise ratio, one can stack the lensing profiles of self-similar galaxies together. The lensing profile of a galaxy correlates directly with its mass. Thus, galaxies sharing a comparable mass exhibit similar  lensing profiles. Consequently, one can combine the lensing profiles of galaxies with analogous mass by stacking them together, yielding a mean profile. The stellar mass distribution of a galaxy can be used to trace its total halo mass \citep{Huang2020, Huang2022}. \cite{Huang2020} has shown that the two-parameter $M_{\rm tot} - M_{\rm inn}$ description provides a more accurate galaxy-halo connection where $\log M_{\rm tot}$ is the total stellar mass of the lensing galaxy and $\log M_{\rm inn}$ is the stellar mass of a lensing galaxy within the 10 kpc radius, both in units of solar mass.To achieve this, it is necessary to select galaxies within a narrow mass bin. \cite{Huang2022} has computed stacked lensing profiles in 12 different stellar mass bins.

We use stacked lensing profiles and corresponding covariance matrices from \cite{Huang2022} in two different mass bins. The mass ranges for first bin (bin 1) are $11.60 \leq \log M_{\rm tot} \leq 11.69$, $10.89 \leq \log M_{\rm inn} \leq 11.28$ and the mass ranges for second bin (bin 2) are $11.87 \leq \log M_{\rm tot} \leq 12.11$, $11.16 \leq \log M_{\rm inn} \leq 11.40$. The mean lens redshifts for bin 1 and bin 2 are $0.36$ and $0.35$, respectively, while the mean source redshift for both is $1.19$. The covariance matrix for the lensing profile of a galaxy summarizes the statistical relationships between measured $\Delta\Sigma$ in different radial bins of a profile. The diagonal elements of the covariance matrix represent the variances of measured $\Delta\Sigma$ in corresponding bins. These stacked profiles and the corresponding covariance matrices are shown in Fig.\ \ref{fig:obser_data}. We choose these two bins because these are representative bins for low and high halo mass groups from the observed HSC data and cover mass range for the TDCOSMO lenses.

\section{Results}
\label{sec:results}
Fig.\ \ref{fig:illustration} illustrates the predicted reduced shear profile of a lens at $z=0.5$ with a halo mass of 14.35 $M_{\odot}$. Both strong and weak lensing phenomena manifest at distinct scales, as depicted in Fig.\ \ref{fig:illustration}. In this paper, our objective is to utilize weak lensing measurements, representing the scale denoted by the light blue shaded region, to constrain MSD in strong lensing mass modeling, measurements corresponding to the light green shaded region. Results obtained from different analyses are described in sections \ref{re:mock_results}, \ref{re:obs_results}, and \ref{sec:forecast_results}. Further refinement of weak lensing measurements within the yellow shaded region in Fig.\ \ref{fig:illustration} holds the potential to enhance the constraint on MSD, a topic that we explore in section \ref{sec:wl_push}.
\subsection{Testing model with the mock data}
\label{re:mock_results}
\begin{figure}
    \includegraphics[width=\linewidth,right]{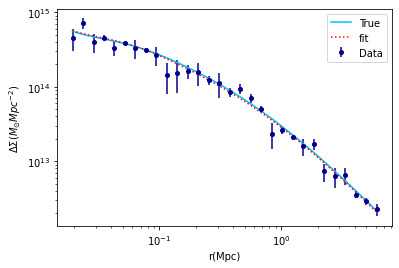}\par
\caption{Mock lensing profile. This profile is generated by using a combination of NFW profile \citep{NFW1996}, Hernquist profile \citep{Hernquist1990}, and superposition of 3 different mass-sheets with varying core radii.}
\label{fig:mock_profile}
\end{figure}

\begin{figure}
    \includegraphics[width=\linewidth,right]{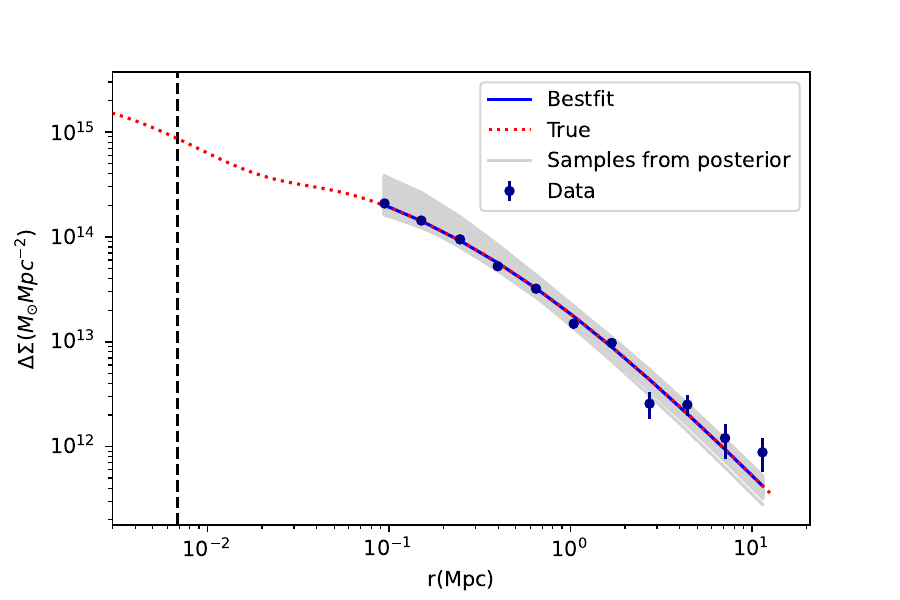}\par
\caption{Mock lensing profile using parameters of observed lensing profile. This profile is generated by using a combination of NFW profile, Hernquist profile, and superposition of 3 different mass-sheets with different core radii. Uncertainties shown in each data points are diagonal elements of covariance matrix of a lensing profile given in Fig.\ \ref{fig:obser_data}. The black dashed vertical line is the Einstein radius of the lens.}
\label{fig:observed_mock_profile}
\end{figure}

\begin{table*}
	\centering
	\small\addtolength{\tabcolsep}{-0pt}
	\small
	\caption{Mock profile fitting results. This Table has a list of recovered mean values for all free parameters in the model used to fit the mock lensing profile shown in Fig.\ \ref{fig:mock_profile}.}
	\label{tab:BFP}
	\begin{threeparttable}
	\begin{tabular}{lccccccc} 
		\hline
		Convention & $\lambda_1 $ & $\lambda_2$  & $\lambda_3$ & $R_{c, 1}$ & $R_{c, 2}$ & $R_{c, 3}$ & $\lambda_{\rm tot}$\\
		\hline
	    Core radii fixed & $0.68^{+0.22}_{-0.42}$ & $1.09^{+0.83}_{-0.38}$ & $1.05^{+0.20}_{-0.39}$  & 7 & 9 & 11 & $0.816^{+0.038}_{0.030}$\\
		\hline
		Core radii free & $0.89^{+0.34}_{-0.43}$ & $0.93^{+0.39}_{-0.46}$ & $0.97^{+0.36}_{-0.44}$  & $7.50^{+2.10_{-4.30}}$ & $8.00^{+2.80}_{-4.00}$ & $8.30^{+3.20}_{-3.80}$ & $0.786^{+0.074}_{0.040}$\\
		\hline
	\end{tabular}
    \end{threeparttable}
\end{table*}

\begin{table*}
	\centering
	\small\addtolength{\tabcolsep}{-0pt}
	\small
	\caption{Results obtained by fitting lensing profile along with observed covariance matrix. This Table has list of recovered mean values for all free parameters in the model used to fit the lensing profile shown in Fig.\ \ref{fig:observed_mock_profile}.}
	\label{tab:BFP_observed}
	\begin{threeparttable}
	\begin{tabular}{lccccccccc} 
		\hline
		Bin & Convention & $\lambda_1 $ & $\lambda_2$  & $\lambda_3$ & $R_{c, 1}$ & $R_{c, 2}$ & $R_{c, 3}$ & $\lambda_{\rm tot}$ & Precision on $\lambda_{\rm tot}$\\
		\hline
	    Bin 1 & Core radii fixed & $0.74^{+0.35}_{-0.35}$ & $0.98^{+0.57}_{-0.57}$ & $1.07^{+0.27}_{-0.27}$  & 7 & 9 & 11 & $0.793 \pm 0.077$ & 9.71$\%$\\
		&Core radii free & $0.88^{+0.42}_{-0.50}$ & $0.89^{+0.41}_{-0.55}$ & $0.97^{+0.43}_{-0.43}$  & $7.40^{3.00}_{-4.30}$ & $7.30^{3.00}_{4.40}$ & $8.00^{+3.80}_{4.20}$ & $0.75^{0.12}_{-0.084}$ & 13.60$\%$\\
		\hline
        Bin 2 & Core radii fixed & $0.65^{+0.39}_{-0.27}$ & $0.93^{+0.35}_{-0.90}$ & $1.15^{+0.42}_{-0.27}$  & 7 & 9 & 11 & $0.728 \pm 0.075$ & 10.30$\%$\\
        & Core radii free & $0.88^{+0.43}_{-0.53}$ & $0.91^{+0.46}_{-0.46}$ & $0.96\pm0.46$  & $7.9^{+3.7}_{-4.2}$ & $8.10\pm 3.60$ & $8.30^{+4.60}_{-3.60}$  & $0.74\pm 0.08$ & 10.81$\%$\\
		\hline
	\end{tabular}
    \end{threeparttable}
\end{table*}

\begin{table*}
	\centering
	\small\addtolength{\tabcolsep}{-0pt}
	\small
	\caption{Results obtained by fitting lensing profile along with observed covariance matrix re-scaled to the larger number of samples. This Table has list of recovered mean values for all free parameters in the model used to fit the lensing profile shown in top panel of left hand side of Fig.\ \ref{fig:observed_mock_profile}.}
	\label{tab:BFP_observed_multiplication}
	\begin{threeparttable}
	\begin{tabular}{lccccccccc} 
		\hline
		Bin & sample number & $\lambda_1 $ & $\lambda_2$  & $\lambda_3$ & $R_{c, 1}$ & $R_{c, 2}$ & $R_{c, 3}$ & $\lambda_{\rm tot}$ & Precision on $\lambda_{\rm tot}$\\
		\hline
	    Bin 1 & $2\times545$ & $0.75^{+0.35}_{-0.35}$ & $0.98^{+0.57}_{-0.57}$ & $1.06^{+0.27}_{-0.27}$  & 7 & 9 & 11 & $0.801^{+0.061}_{-0.055}$ & 7.24$\%$\\
        & $10\times545$ & $0.80^{+0.48}_{-0.27}$ & $0.95^{+0.46}_{-0.79}$ & $1.07^{+0.37}_{-0.21}$  & 7 & 9 & 11 & $0.816^{0.057}_{-0.040}$ & 5.94$\%$\\
        & $20\times545$ & $0.81^{+0.40}_{-0.30}$ & $0.94^{+0.51}_{-0.65}$ & $1.07^{+0.31}_{-0.24}$  & 7 & 9 & 11 & $0.821^{0.054}_{-0.034}$ & 5.42$\%$\\
        & $140\times545$ & $0.82\pm0.19$ & $0.93\pm0.32$ & $1.07\pm 0.15$  & 7 & 9 & 11 & $0.826^{0.029}_{-0.021}$ & 3.03$\%$\\
        & $500\times545$ & $0.81\pm0.11$ & $0.96\pm0.17$ & $1.06\pm 0.082$  & 7 & 9 & 11 & $0.825^{0.019}_{-0.008}$ & 1.64$\%$\\
		\hline
        Bin 2 & $2\times78$ & $0.72^{+0.40}_{-0.22}$ & $0.90^{+0.39}_{-0.88}$ & $1.13^{+0.44}_{-0.23}$  & 7 & 9 & 11 & $0.757^{+0.053}_{-0.047}$ & 6.61$\%$\\
        & $10\times78$ & $0.84^{+0.36}_{-0.16}$ & $0.78^{+0.25}_{-0.75}$ & $1.17^{+0.39}_{-0.16}$  & 7 & 9 & 11 & $0.800^{0.034}_{-0.022}$ & 3.50$\%$\\
        & $20\times78$ & $0.87^{+0.32}_{-0.15}$ & $0.76^{+0.30}_{-0.64}$ & $1.18^{+0.34}_{-0.16}$  & 7 & 9 & 11 & $0.810^{0.028}_{-0.016}$ & 2.72$\%$\\
        & $140\times78$ & $0.86^{+0.12}_{-0.12}$ & $0.83^{+0.23}_{-0.23}$ & $1.13^{+0.13}_{-0.13}$  & 7 & 9 & 11 & $0.822^{0.014}_{-0.006}$ & 1.34$\%$\\
        & $500\times78$ & $0.829^{+0.066}_{-0.066}$ & $0.90^{+0.12}_{-0.12}$ & $1.093^{+0.068}_{-0.068}$  & 7 & 9 & 11 & $0.824^{0.010}_{-0.001}$ & 0.67$\%$\\
		\hline
	\end{tabular}
    \end{threeparttable}
\end{table*}

We test our model given in Section$\sim$ \ref{sec:proposed_model} to see its ability to recover free parameters from the likelihood analysis. For this, we generate a mock lensing profile using a  combination of NFW profiles, Hernquist profiles, and superposition of three different mass sheets with different core radii. This profile is shown in Fig.\ \ref{fig:mock_profile}. To generate this profile, we use a lens redshift $(z_l) = 0.5$ and a source redshift $(z_s) = 1.5$. Halo mass $(M_h)$ and concentration $(c)$ for the NFW profile are $10^{14} M_{\odot}$ and 4, respectively. For the Hernquist profile, stellar mass $M_s = 10^{11}$ and scale radius $r_s = 5$ $\rm kpc$ have been used. For three different mass sheets, we use $\lambda_1, \lambda_2, \lambda_3$ equal to 0.8, 0.98, and 1.05 respectively, and $R_{c, 1}, R_{c, 2}, R_{c, 3}$ equal to 7, 9, and 11 arcsec respectively. The Einstein radius of this lens is 0.67 arcsec and the true value of $\lambda_{\rm tot}$ is $0.83$. For a likelihood analysis, we need a $1\sigma$ uncertainty on each $\Delta\Sigma$ measurement in the profile. This is a mock profile so it does not have uncertainties on each data point. Therefore, we have assigned $20\%$ uncertainty in this mock lensing profile i.e.\ $\sigma_{\Delta\Sigma} = 0.2 \Delta\Sigma$ and added random Gaussian noise. One can assign any other reasonable amount of uncertainty based on their own requirement.

\begin{figure*}
\begin{multicols}{2}
    \includegraphics[width=\linewidth]{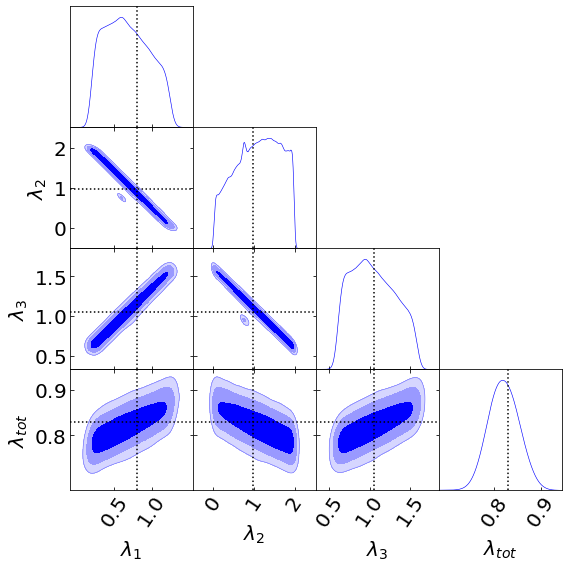}\par
    \includegraphics[width=\linewidth]{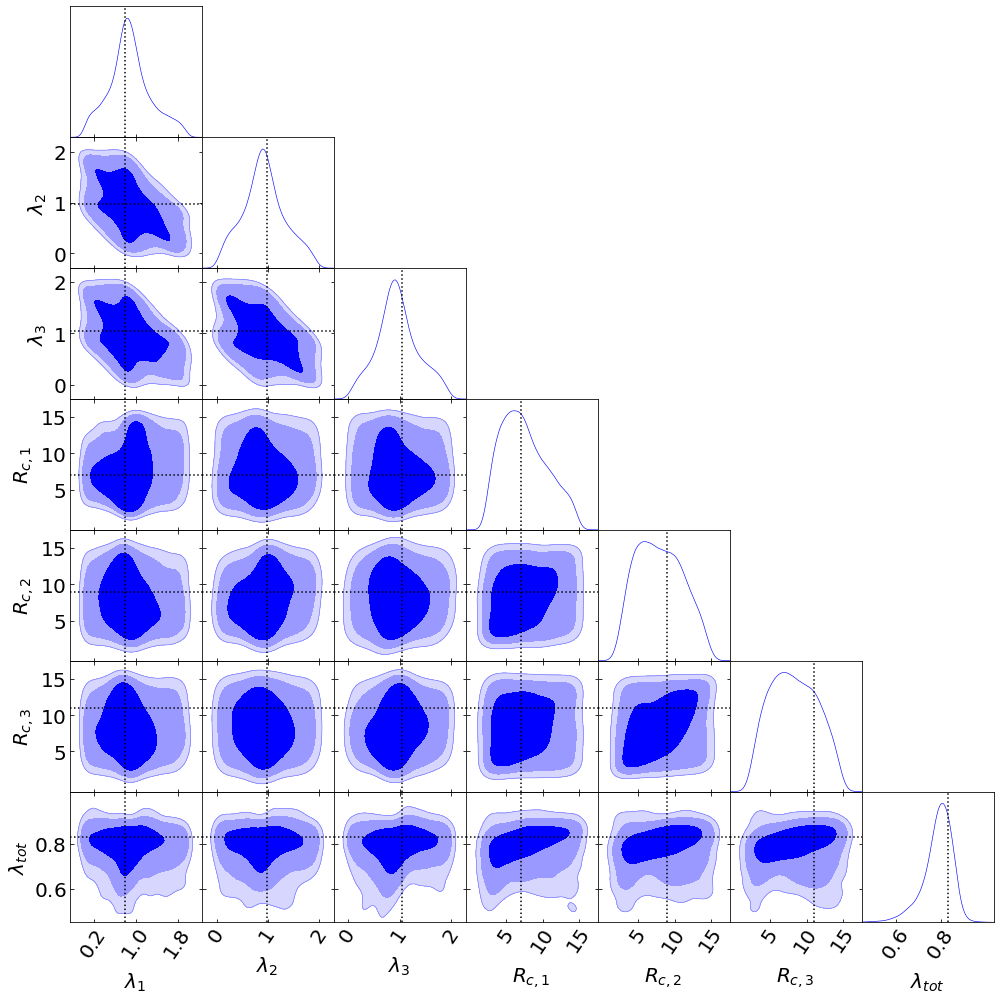}\par
\end{multicols}
\caption{One-dimensional likelihood distributions and two-dimensional contours at 1$\sigma$, 2$\sigma$, and 3$\sigma$ confidence levels for all free parameters using pure mock profile. Left panel corresponds to the core radii fixed case and the right panel corresponds to the core radii free case. Black dotted lines in all plots represent true values of corresponding free parameters. In both cases, true values of all the free parameters are well recovered through likelihood analysis. Posterior in these plots show that constraints on each MST factors $\lambda_1$, $\lambda_2$, and $\lambda_3$ are loose, but $\lambda_{tot}$ is relatively well constrained.}
\label{fig:mock_fitting}
\end{figure*}

We fit this profile using the MCMC simulation and results are shown in Table \ref{tab:BFP} and corresponding plots are shown in Fig.\ \ref{fig:mock_fitting}. From numbers listed in Table \ref{tab:BFP}, we see that MCMC fittings are able to recover each individual mass-sheet transformation factors $\lambda_i$ with large error bars in both fixed core radii and free core radii cases. However, superimposed or total mass-sheet transformation factor $(\lambda_{\rm tot})$ is well constrained in comparison to individual $\lambda_i$ constraints. In case of fixed core radii, the $\lambda_{\rm tot}$ is constrained with $4.17 \%$ precision while in case of free core radii, it is constrained with $7.25 \%$ precision. In this likelihood analysis of a mock profile, the precision of constraints on the $\lambda_{\rm tot}$ depends on the choice of uncertainties applied to the mock lensing profile. As a result, the precision associated with these $\lambda_{\rm tot}$ constraints lacks absolute significance. However, in both cases, the true value of $\lambda_{\rm tot}$ resides within the posterior distribution of $\lambda_{\rm tot}$. This shows that weak lensing has the capability to measure the total mass sheet transformation ($\lambda_{\rm tot}$) in our model. Notably, $\lambda_{\rm tot}$ stands as the sole quantity necessary for the $H_0$ constraint.

\subsection{Constrain in the MST of a mock lensing profile using covariance matrix of the observational data}
\label{re:obs_results}
We use stacked galaxy-galaxy lensing profiles from \cite{Huang2022} which are shown in Fig.\ \ref{fig:obser_data}. We use covariance matrices of these profiles to fit mock lensing profiles. We use all the parameters (halo mass, stellar mass, redshifts, mean radial distance in each radial bin) of these lensing profiles and MST parameters described in Section \ref{re:mock_results} to generate mock lensing profiles and fit these mock lensing profiles using covariance matrices of observed lensing profiles. For simplicity, we fix the halo mass and stellar mass in our model during the fitting. However, in a practical application, one can fit the weak lensing profile using a parametric model such as an elliptical power law along with mass sheets which is jointly fit with the strong lensing data. We refer to Appendix \ref{sec:appendix} for this more realistic approach and state that the findings in regard to constraints on the MST from weak lensing measurements remain consistent with this section. Mock lensing profiles are generated using eq.\ \ref{eq:dsigma}. The Einstein radius of mock lenses of bin 1 and bin 2 are 1.36 arcsec and 3.14 arcsec, respectively. In both cases, the true value of $\lambda_{\rm tot}$ is $0.83$. A mock lensing profile generated using parameters of observed lensing profile of bin 1 (upper left panel of Fig.\ \ref{fig:obser_data}) is shown in Fig.\ \ref{fig:observed_mock_profile}. In this figure, one can also see how well the best-fit profile matches the true mock lensing profile.

We fit these profiles using MCMC simulation and the results are shown in Table \ref{tab:BFP_observed} and corresponding plots are shown in gray color in Fig.\ \ref{fig:obs_fitting_multiplication}. From these numbers and figures, we can see that MCMC fitting of mock lensing profile using covariance matrix of observed lensing profile is able to recover all the true parameters of mock lensing profile in both core radii fixed and free cases with wide error bars. However, the total mass sheet transformation factor $(\lambda_{\rm tot})$ is well constrained in both cases. For bin 1, in case of fixed core radii, the $\lambda_{\rm tot}$ is constrained with $9.71\%$ precision while in case of free core radii, it is constrained with $13.6\%$ precision. For bin-2, in case of fixed core radii, the $\lambda_{\rm tot}$ is constrained with $10.30\%$ precision while in case of free core radii, it is constrained with $10.81\%$ precision. These results demonstrate that the covariance matrices of observed lensing profiles, obtained from the HSC data within specific mass bins, can constrain the MST with approximately $10\%$ precision and for both bins, as expected from Section$\sim$ \ref{re:mock_results} fitting with fixed core radii provides relatively better constraints on $\lambda_{\rm tot}$.

In the generation of mock lensing profiles, we used NFW and Hernquist profiles to predict the convergence $(\kappa(\vec{\theta}))$ and also applied these models during the fitting process. This leads to the question of whether one
can use different sets of model for $\kappa(\vec{\theta})$ during the fitting. 
To address this question, we generate a mock lensing profile using the NFW + Hernquist model and fit this profile using the elliptical power law (EPL) model with a different number of mass sheets. The results from this analysis demonstrate that in fitting, one can use a model different from the one used in mock data generation but must include a sufficient number of mass sheets so that the strong lensing potential from the bestfit model matches the true potential. A description of these results and the corresponding plots are given in Appendix \ref{sec:appendix}.

\subsection{Forecast for an LSST-like survey}
\label{sec:forecast_results}
The Legacy Survey of Space and Time (LSST) will survey about 20,000 square degrees of the Southern Hemisphere sky, providing extensive weak lensing data \citep{LSST2019}. Euclid, covering 15,000 square degrees, focuses on cosmological studies and dark energy using gravitational lensing, also contributing to weak lensing data \citep{Euclid2022}. In contrast, HSC data used in Section \ref{re:obs_results} covers only 137 $\rm deg^2$, making it approximately 145 times smaller than LSST and 110 times smaller than Euclid. If we attempt to constrain $\lambda_{\rm tot}$ using all available HSC galaxy-galaxy lensing data, we achieve $\sim 10\%$ precision in the constraint on $\lambda_{\rm tot}$ which is not sufficient for the precise measurement of the Hubble constant $(H_0)$. However, the wider coverage of LSST and Euclid surveys can provide a significantly large amount of weak lensing data compared to the HSC survey and could lead to reasonably tight constraints on $\lambda_{\rm tot}$. To explore this further, we increased the sample size of each stacked lensing profile given in Fig.\ \ref{fig:obser_data} by different factors to observe the corresponding changes in $\lambda_{\rm tot}$ constraints with sample size and results are described below.

In this section, we analysed the same mock lensing profiles described in Section \ref{re:obs_results} but with rescaled covariance matrices. We increased the number of samples in the stacked lensing profiles by 2, 10, 20, 140, and 500 times and performed analyses with each case and results are listed in Table \ref{tab:BFP_observed_multiplication}.\footnote{The covariance matrix depends on sample size. If sample size in a stacked lensing profile increases by factor of $x$, the corresponding covariance matrix shrinks by the factor of $1/x$. Therefore, we divide the covariance matrix by 2, 10, 20, 140, and 500 and consider that is equivalent to increasing sample size by corresponding factors.
} For both profiles, one-dimensional likelihood distribution and two-dimensional contours for each free parameters for increased sample size by 140 times (a factor achievable by LSST survey) are shown in Fig.\ \ref{fig:obs_fitting_multiplication} in blue color. From Table \ref{tab:BFP_observed_multiplication}, if we double the current sample size of the HSC data, we can achieve $6.61 - 7.24 \%$ precision on $\lambda_{\rm tot}$ constraint. Similarly, if we increase the sample size by 10, 20, and 140 times, we can achieve $3.50 - 5.94 \%$, $2.72 - 5.42 \%$, and $1.34 - 3.03 \%$ precision on $\lambda_{\rm tot}$ respectively. On the other hand, to achieve $0.67-1.64 \%$ precision we need to increase our current HSC galaxy-galaxy sample by 500 times.

On the basis of the survey area of LSST, we can expect to achieve at least 140 times the current galaxy-galaxy lensing samples of the HSC data. Therefore, we expect to achieve $\sim 1.34 - 3.03 \%$ precision on $\lambda_{\rm tot}$ constraint using galaxy-galaxy lensing samples from the LSST data. Similarly, the Euclid survey could also provide similar constraints on $\lambda_{\rm tot}$. On the other hand, 500 times the current galaxy-galaxy lensing sample size of HSC which leads to $\sim 1\%$ precision on the $\lambda_{\rm tot}$ constraint is not achievable from the LSST survey and need a larger survey than LSST or combination of data from different survey. 

\subsection{Impact of smaller angular scale weak lensing measurements on the MSD constraint}
\label{sec:wl_push}
\begin{figure*}
\begin{multicols}{2}
    \includegraphics[width=\linewidth]{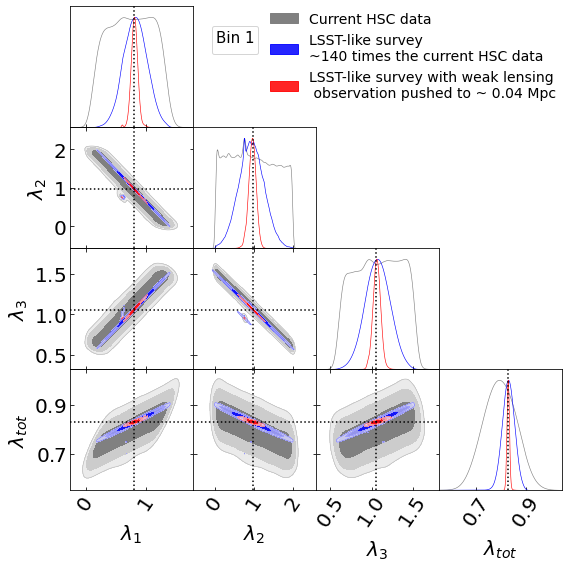}\par
    \includegraphics[width=\linewidth]{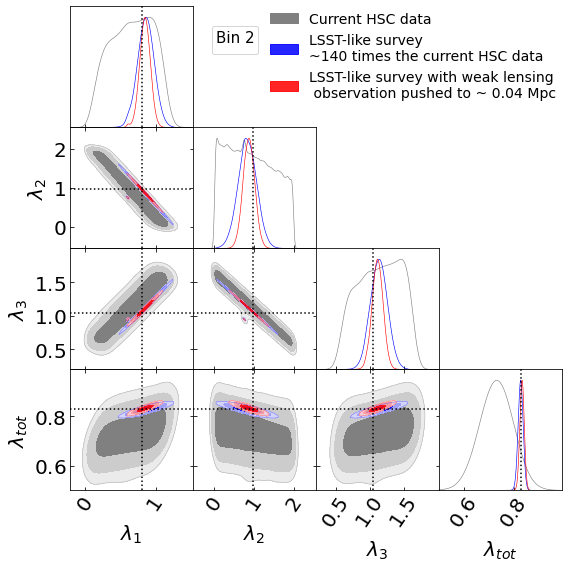}\par
\end{multicols}
\caption{One-dimensional likelihood distributions and two-dimensional contours at 1$\sigma$, 2$\sigma$, and 3$\sigma$ confidence levels for all free parameters using exact and re-scaled covariance matrices from  Fig.\ \ref{fig:obser_data} (re-scaled by factor of 140). The left panel corresponds to bin 1 and the right panel corresponds to bin 2. In both bins, the current HSC data, shown in gray color, provides $\sim 10\%$ precision on $\rm \lambda_{tot}$. A LSST-like survey, shown in blue color, which can provide $\sim 140$ times the current HSC galaxy-galaxy lensing sample, will lead to $\sim 1-3 \%$ precision on $\rm \lambda_{\rm tot}$ constraint. On the other hand, if we extend weak lensing measurements of a LSST-like survey to lower scale of $\sim 0.04$ $\rm Mpc$, this will lead to $\sim 1\%$ precision on $\rm \lambda_{tot}$ constraint, depicted in red color for both bins. Dotted black lines in all sub-panels represent the true value of corresponding free parameters.
}
\label{fig:obs_fitting_multiplication}
\end{figure*}

Weak lensing profiles in Fig. \ref{fig:obser_data} are limited to a $R > 100$ $\rm kpc$ scale, typical for weak lensing observations. To explore the impact of including smaller scales (< 100 $\rm kpc$), we extend the data and associated covariance matrices to $\sim 40$ $\rm kpc$ by adding two more radial bins. At scales less than 100 $\rm kpc$, shape noise ($\sigma_{\rm shape}$) dominates measurement error. We approximate the diagonal terms of covariance matrix for the added bins by shape noise, assuming that the corresponding off-diagonal terms are zero. The average shape noise for HSC weak lensing is $\sim 0.4$, with a weighted galaxy density of $21.8$ $\text{ arcmin}^{-2}$ \citep{Mandelbaumetal2017}. Utilizing this noise and density, we compute shear uncertainty $(\sigma_{\rm shear})$ using $\sigma_{\rm shear} = \sigma_{\rm shape}/(2 \times n_g)^{0.5}$, where $n_g$ represents the galaxy density. We generated mock lensing profiles as described in \ref{re:obs_results} including two more radial bins in lower scale.  We performed a likelihood analysis of these profiles, using extended covariance matrices rescaled to a sample size of 140 times the current HSC galaxy-galaxy lensing sample. These extended covariance matrices achieve $\sim 1\%$ precision on $\lambda_{\rm tot}$ constraint. The results are depicted in Fig. \ref{fig:obs_fitting_multiplication} in red color, indicating that weak lensing profiles with measurements below a 100 $\rm kpc$ scale can lead to better constraints on $\lambda_{\rm tot}$.

\section{Discussion}
\label{sec:discussion}
The observed galaxy-galaxy lensing profile faces a limitation in reaching well inside the lens mass distribution, as illustrated in Figs.\ \ref{fig:illustration} and \ref{fig:observed_mock_profile}. Notably, the weak lensing data (depicted by dark blue dots) extend only up to 0.1 $\rm Mpc$,  while the Einstein radius, at 0.004 $\rm Mpc$ (the black dotted vertical line), remains at a smaller scale. This discrepancy introduces a level of freedom between the Einstein radius of the lens galaxy and the innermost point of the observed weak lensing profile, contributing to uncertainties in the MST constraint. To mitigate this challenge and enhance the precision of the MST constraints, it becomes imperative to measure the galaxy-galaxy lensing profile to smaller angular separation to the deflector galaxy. Such small angular scale measurements hold the potential to yield more precise constraints on the MST by probing the finer details of the lens mass distribution which is indicated by result in section \ref{sec:wl_push}.

At lower scales the aperture for the source galaxies are relatively small. Therefore, the source galaxy catalog could have many overlapping pairs of galaxies. This could result in biased galaxy shape measurements and biased shape measurements directly lead to biased lensing profiles \citep{Kobayashi2015}. One can exclude such overlapping galaxy pairs from the source catalog and can avoid this issue. However, this will reduce the source number density and result in a wider error bar on weak lensing measurements. On the other hand, at sufficiently smaller scales, the presence of lens light may introduce contamination in photometry and shape measurement of a source galaxy \citep{Kobayashi2015}. Biased photometry leads to the biased photometric source redshift. A weak lensing profile depends on the $\Sigma_{\rm crit}$ so biased photometric source redshift leads to the biased weak lensing profile. Therefore, either one can exclude scales where this effect starts to impact the weak lensing profile or can model both the lens and source light components accurately which will increase the complexity of the analyses. Mitigating these effects requires advanced techniques such as  sophisticated modeling, and statistical approaches to ensure the reliability of shear measurements.

Stacked weak lensing profiles are one of the key ingredients for our methodology. The selection of the deflector galaxies for the stacked lensing profile must match the deflector galaxies of the strong lenses with time-delay measurements. 

A primary property that needs to be matched is the halo mass of the deflector. The halo mass is not directly observable. Therefore, one needs to use a halo mass proxy. The biased tracer of halo mass can lead to the biased lensing profile and can impact our method. In this work, we used the combination of inner and outer stellar mass as a proxy of halo mass \citep{Huang2022} and we select subsamples of galaxies in narrow ranges within these two quantities.
An additional source of possible systematic bias is the strong lensing selection effect. Strong lenses are more likely to reside in more concentrated projected mass distributions \citep[e.g.,][]{Sonnenfeld:2023}. The key aspect is that within a narrowly selected weak lensing deflector sample, the strong lensing selection bias needs to be understood. Forward simulating the strong lensing quantities allow to quantify this selection effect. Further studies in the context of current and future surveys in regard to the relative selection effect between strong and weak lensing deflector samples are required. 

Galaxies used in the stacked lensing profile can exhibit some scatter in the MST constraints.  During cosmological inference, this scatter can be modeled and its value can be inferred on the population level, as e.g. done by \cite{Birrer2020}. 
The scatter should be consistent with the residuals between the time-delay measurements and the predictions. The value of the scatter depends on the similarity of the stacked lenses. Therefore, if we can accurately select self-similar galaxies, the scatter in $\lambda_{\rm tot}$ should be be small enough, which should not limit the cosmological inference.

In this paper, we have scaled the HSC galaxy-galaxy lensing sample to LSST based on the survey area while maintaining the narrowly defined halo mass selection by \cite{Huang2022}. Another factor that can impact the sample size is the redshift range of the deflector galaxies. Galaxy sample of \cite{Huang2022} span the redshift range $0.19 < z < 0.51$. On the other hand, TDCOSMO lenses span a wider redshift range of $0.30 \leq z \leq 0.75$ \citep{Birrer2020}. The extension of a galaxy-galaxy lensing sample in the range $0.51 \leq z \leq 0.75$ can further increase the sample size of LSST galaxy-galaxy lensing and improve the precision of the computed lensing profile, thereby further enhancing the constraints on the MSD.
In addition, while we have derived constraints and forecast for a single halo mass bin, the time-delay lenses are likely to occupy a wider mass range, and hence several mass bins may contribute to our signal, which further can enhance the precision on the combined $H0$ measurement.

\section{Conclusions}
\label{sec:conclusion}
In this paper, we modeled an approximate MST as a superposition of several cored profiles. The robustness of the proposed model is assessed through comprehensive testing, employing mock lensing profiles generated with a combination of NFW and Hernquist profiles, along with three different mass sheets with distinct core radii. The proposed model is flexible enough to generalize the mass density profile subjected to approximate MSTs. The MCMC simulations demonstrate the model's capability to recover free parameters associated with MSTs from likelihood analysis, particularly emphasizing the well-constrained superimposed mass sheet transformation factor ($\lambda_{\rm tot}$), which holds significance for the $H_0$ constraint. 

We proposed and investigated the possibility of using galaxy-galaxy weak lensing shear measurements to constrain the MSD. We constrained the MST using covariance matrices derived from observed galaxy-galaxy lensing profiles. The current HSC lensing observations are able to constrain the MST with $\sim 10\%$ precision. In anticipation of the upcoming wide and deep surveys, we investigated their potential to achieve the more accurate constraint on the MST. Given that the data utilized in this paper from the HSC survey covers $137$ $\rm deg^2$, the future LSST survey, with a size of $20,000$ $\rm deg^2$, can provide observations equivalent to $\sim 140$ times the current HSC galaxy-galaxy lensing sample size when scaled accordingly. Therefore, LSST survey will provide a reasonable amount of galaxy-galaxy lensing observations that will be sufficient to achieve $\sim 1-3 \%$ precision on the MST  constraint. If we can push lensing observations to the lower scale around $\sim 0.04$ $\rm Mpc$, LSST survey will provide $\sim 1\%$ precision on the MST constraint. On the other hand, 500 times the current galaxy-galaxy sample of HSC can constrain the MST with $\sim 1\%$ precision. This large galaxy-galaxy lensing sample size could be achieved from the combined data of LSST, Euclid, and Roman. A higher number density of source galaxies in a survey generally leads to a more accurate and precise measurement of the excess surface mass density because a larger sample size provides better statistics, reducing the impact of random fluctuations and observational noise. The Euclid survey could have a higher density of source galaxies relative to the ground based survey. So, we could possibly compute a precise lensing signal and this could also further enhance the constraint on $\lambda_{\rm tot}$.

We believe that a  promising method to break the mass sheet degeneracy in strong lensing mass modeling is the utilization of weak lensing shear measurements, which circumvents systematics linked with secondary assumptions in density profile modeling. This method allows us to use already obtained survey data to break the MSD for the observed strong lenses which will somehow complement other time and resource extensive methods and help us to speed up our strong lensing analyses. We anticipate that LSST will observe more than 100 strongly lensed quasars and nearly 10 lensed SNIa suitable for TDC each year.  This data will be sufficient to reduce the statistical uncertainty in the measurement of $H_0$ to a sub-percent level, with the error budget of the $H_0$ value primarily dominated by the uncertainty in the constraint of $\lambda_{\rm tot}$. Therefore, our forecast for LSST like survey with our proposed method demonstrates that a $1-3\%$ measurement on $H_0$ combining time delays, strong lens modeling, and galaxy-galaxy weak lensing measurements is feasible. Furthermore, if we extend weak lensing measurements of LSST-like surveys to a lower scale of $\sim 0.04$ $\rm Mpc$, we will be able to achieve $\sim 1 \%$ precision on the MST constraint and hence $\sim 1\%$ precision on the $H_0$ measurement.

\section{ACKNOWLEDGEMENTS}
This work is supported by the Stony Brook University, Department of Physics and Astronomy. We thank the organizers of the 2023 KICP workshop: "lensing at different scales: strong, weak, and synergies between the two".

\section*{Data availability}
The observed galaxy-galaxy lensing measurements used in this paper are publicly available in \cite{Huang2022}.



\bibliographystyle{mnras}
\bibliography{mybibfile}



\begin{appendix}
\section{Comparison between the fitting accuracy of EPL and EPL + MST models.}
\label{sec:appendix}
\begin{figure*}
\begin{multicols}{2}
    \includegraphics[width=\linewidth, height=7cm]{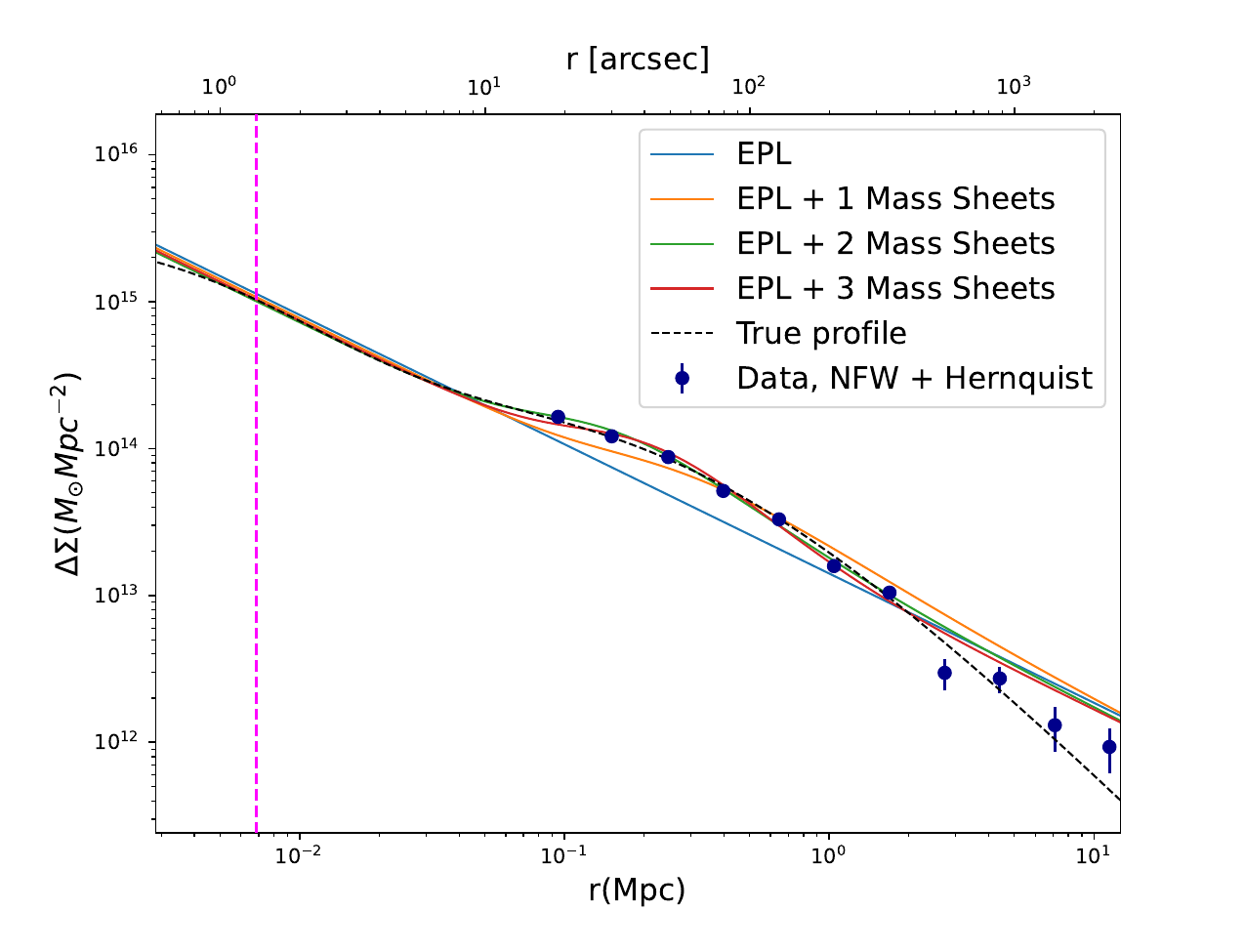}\par
    \includegraphics[width=\linewidth, height=7cm]{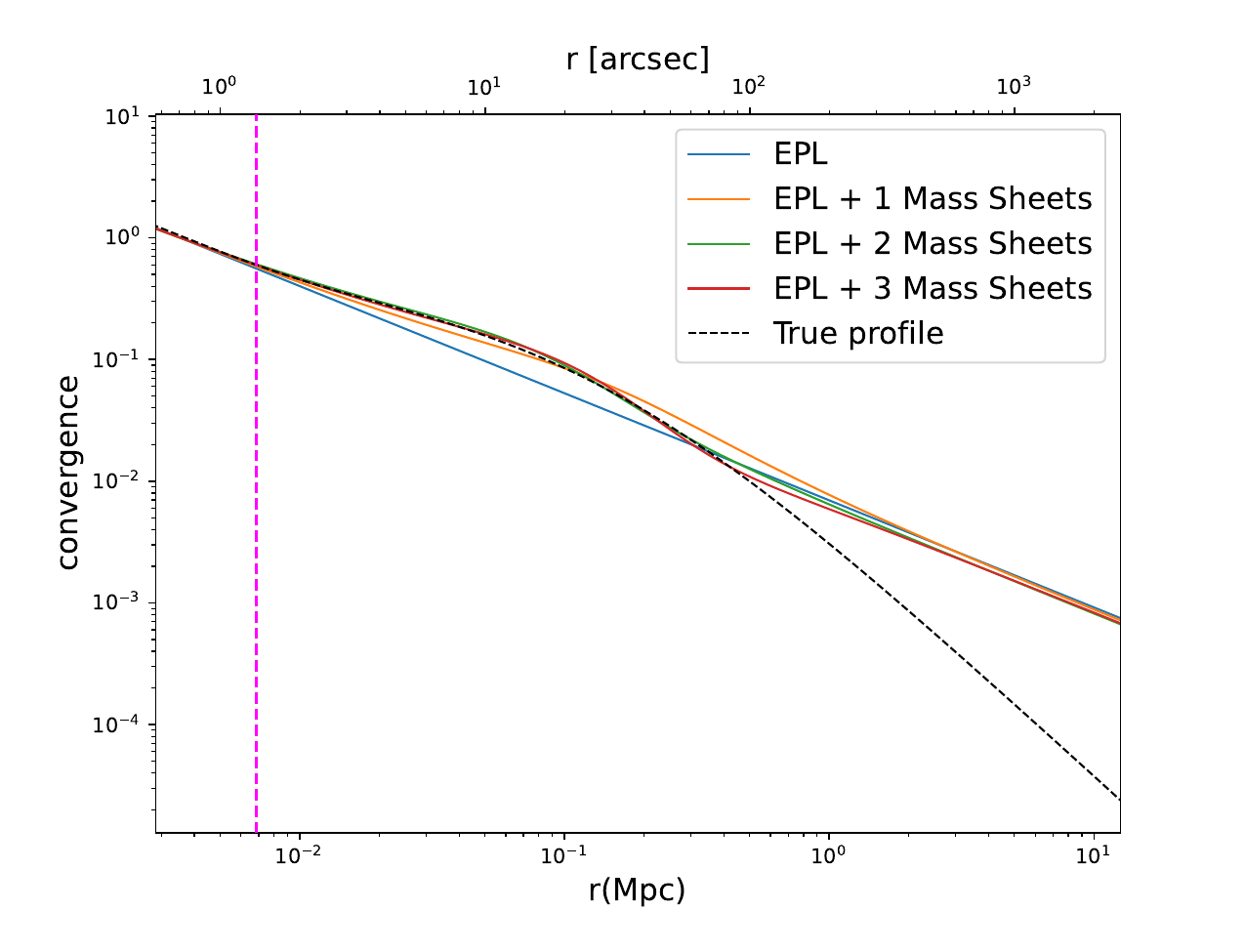}\par
\end{multicols}
\caption{Left panel: Comparison between the true weak lensing profile (dotted black) and the best-fit weak lensing profile using EPL (blue) and EPL with different numbers of mass sheets (orange, green, red) models. The analysis is performed using the covariance matrix of the lensing profile of bin 1. The dark blue points with error bars represent the mock data used in the analysis, and this profile is generated using the NFW + Hernquist model. Including more mass sheets fits the data better, and the best-fit model is closer to the true profile in both the strong and weak lensing regimes. Right panel: The same comparison as the left panel but in terms of convergence profiles. Under each MST case, the Einstein radius of the profile is conserved. The differences between the true convergence and the best-fit convergence at the Einstein radius for EPL, EPL + 1 mass sheet, EPL + 2 mass sheets, and EPL + 3 mass sheets are $6.54\%$, $2.88\%$, $1.57\%$, and $0.19\%$, respectively. Therefore, the EPL model alone cannot recover the true convergence, but reasonably precise convergence can be recovered by including a higher number of mass sheets.}
\label{fig:NFW_EPL}
\end{figure*}

In time-delay cosmography, the measurement of $H_0$ scales with the value of convergence at Einstein radius ($\kappa_{E}$) i.e.\ $H_{0} \propto (1 - \kappa_{E})$. Therefore, if the model used to fit the lensing profile provides biased $\kappa_{E}$, this will lead to the biased $H_0$ measurement. So, can one recover 
accurate value of $\kappa_{E}$ if one fits the lensing profiles using a different set of mass models than used in the mock lensing profile generation? To address this question, we generated a mock lensing profile using NFW + Hernquist model (dark blue points in the left panel of Fig.\ \ref{fig:NFW_EPL}). Parameters for the NFW and Hernquist model are the same with those given in profile of bin 1 (see top left panel of Fig.\ \ref{fig:obser_data} and section \ref{re:obs_results}). We fit this profile using the elliptical power law (EPL) profile, satisfying the same strong lensing constraints as the true profile and including a different number of mass sheets. Fitting results are shown in Fig.\ \ref{fig:NFW_EPL}. Left panel of the Fig.\ \ref{fig:NFW_EPL} shows the comparison between the true lensing profile and the best fit models with the different numbers of mass sheets. From this plot, one can see that the model with three mass sheets can fit the data better than the model with less number of mass sheets and bestfit model in this case is more close to the true profile at both strong and weak lensing regime. Right panel shows the comparison between the true convergence profile and the bestfit convergence profiles with the EPL and EPL plus one, two, and three mass sheets. The differences between the true convergence and the best-fit convergence at the Einstein radius for EPL, EPL + 1 mass sheet, EPL + 2 mass sheets, and EPL + 3 mass sheets are $6.54\%$, $2.88\%$, $1.57\%$, and $0.19\%$, respectively. This clearly shows that the $\kappa_{E}$ can be recovered with a sufficient accuracy using model (different than true model) with the sufficient number of mass sheets. Therefore, including higher number of mass sheets acts as a correction to the EPL profile and helps to recover $\kappa_{E}$ and hence leads to the accurate measurement of $H_0$.
\end{appendix}


\bsp	
\label{lastpage}
\end{document}